\documentclass[aps,prc,showpacs,amsmath,amssymb]{revtex4}
\usepackage{epsfig}

\newcommand{\acal}{{\cal A}}

\begin{document}

\title{In-Medium Modifications of the $\boldsymbol{S_{11}(1535)}$ Resonance 
     and Eta Photoproduction}

\author{J. Lehr, M. Post and U. Mosel}

\affiliation{Institut f\"ur Theoretische Physik, Universit\"at
Giessen\\ D-35392 Giessen, Germany}

\date{\today}

\begin{abstract}
We investigate the influence of possible in-medium modifications of the
$S_{11}(1535)$ in eta photoproduction in nuclei.
Besides Fermi motion, Pauli blocking and binding effects also collisional
broadening is accounted for. The in-medium width is obtained from a
realistic resonance-hole model. Results on eta photoproduction are
obtained from a semi-classical BUU transport model and compared with
data. We find that calculations including a momentum dependent nucleon
and resonance potential are in agreement with the recent KEK data.
In contrast, collisional broadening has only little influence.
\end{abstract}
\maketitle

\section{Introduction}

Photon induced reactions on nuclei provide a promising tool for the
investigation of in-medium properties of hadrons. However, 
the interpretation of data is often difficult,
because it is hard to disentangle contributions from different 
resonances and/or different decay channels as e.g. in the case of pion 
production in the second resonance region.
In this respect the study of the $S_{11}(1535)$ is very interesting because 
of its strong coupling to $N\eta$. 
In the energy regime of $E_\gamma\sim 600-900$ MeV this channel is strongly
dominated by the $S_{11}(1535)$ and therefore $\eta$ photoproduction gives
information almost solely about this resonance.

In the past, different experiments of $\eta$ photoproduction on several
nuclei (C, Al, Ca, Cu, Nb, Pb) were performed. The TAPS group covered the 
energy range from threshold up to photo energies of 800 MeV \cite{roebig_eta},
whereas measurements at KEK \cite{yorita,yamazaki} provided data up 
to 1 GeV.
Other theoretical approaches to the process discussed several 
in-medium effects.
In \cite{yorita} a QMD model was applied. Besides the trivial effects of
Fermi motion and Pauli blocking, the authors found a strong influence due to
collisional reactions of the $S_{11}$.
The authors of \cite{maruyama} were able to describe the KEK data on Carbon 
under the assumption that both scalar and vector potential of the $S_{11}$ 
vanish. 
However, the calculations were performed in nuclear matter and the $\eta$ 
final state interactions were modelled by using a constant absorption factor.
Hence, the findings of this work have to be checked by a more realistic
model.

In this work, we calculate $\eta$ photoproduction on different nuclei using
a semi-classical BUU transport model, which was already successfully applied
to the calculation of a variety of reactions (e.g. heavy-ion reactions
\cite{teis}, photon- and electron-induced reactions 
\cite{effe_pion,lehr_electro}). Within this model, $\eta$ photoproduction 
for energies up to 800 MeV  was already addressed in 
\cite{effe_pion,hombach}. In contrast, we now want to extend our study 
to higher energies covered by the KEK data and to discuss medium modifications
of the $S_{11}$. The $S_{11}$ in-medium width is obtained from self-consistent 
resonance-hole calculations \cite{post}.

We start with a brief review of the BUU model and the implementation
of the $\eta$ and $S_{11}$ dynamics in Sec. \ref{sec:buu_model} followed
by a discussion of in-medium modifications of the $S_{11}$ width in
Sec. \ref{sec:inmed_width}. In Sec. \ref{sec:results} we show our
results in comparison with experimental data.


\section{The BUU Model} \label{sec:buu_model}

For the present studies, we use the BUU model reported in 
\cite{lehr_electro,effe_dilep}. 
The model is based upon a system of coupled BUU equations describing the 
evolution of the spectral distribution function $F_i=\acal_i f_i$ with the
phase space density $f_i$ of  
different particle types $i=N,\pi,S_{11}(1535),\eta,...$:
\begin{equation}
  \label{eq:buu-eq}
  \left({\partial\over\partial t}+\vec\nabla_p H\cdot\vec\nabla_r
  -\vec\nabla_r H\cdot\vec\nabla_p\right)F_i(\vec r,\vec p,\mu;t)
=\Sigma^<_i\acal_i(1-f_i(\vec r,\vec p,\mu;t))-\Sigma^>_i\acal_i 
  f_i(\vec r,\vec p,\mu;t).
\end{equation}
Here  $\acal_i$ is the spectral function and $H$ is a relativistic Hamilton 
function 
\begin{equation}
  \label{eq:hamilton-fkt}
  H=\sqrt{(m+S)^2+p^2}
\end{equation}
with a scalar potential $S$ described in Eq. (\ref{eq:potential3}).

$\Sigma^{\gtrless}_i$ stand for the collision rates of particle species $i$
and describe the gain and the loss of the distribution function at some
'point' $(\vec r,\vec p,\mu,t)$ due to collisional reactions with particles
of the same and other types. Therefore, the BUU equations are coupled
via their right-hand sides. The set of 
equations is solved by a test particle ansatz for each density function $F_i$.
For model details we refer the reader to \cite{lehr_electro,effe_dilep}.

Besides the nucleon, our model contains 29 nucleon resonances with
parameters taken from the analysis of Manley and Saleski \cite{manley} and
all relevant mesonic degrees of freedom ($\pi,\eta,\rho,\omega$).
The $\eta$ couples to the resonances $S_{11}(1535)$,
$S_{11}(1650)$ and $F_{17}(1990)$ with pole-mass branching ratios of
0.43, 0.03 and 0.94.

\begin{figure}
\begin{center}
\includegraphics[width=11cm]{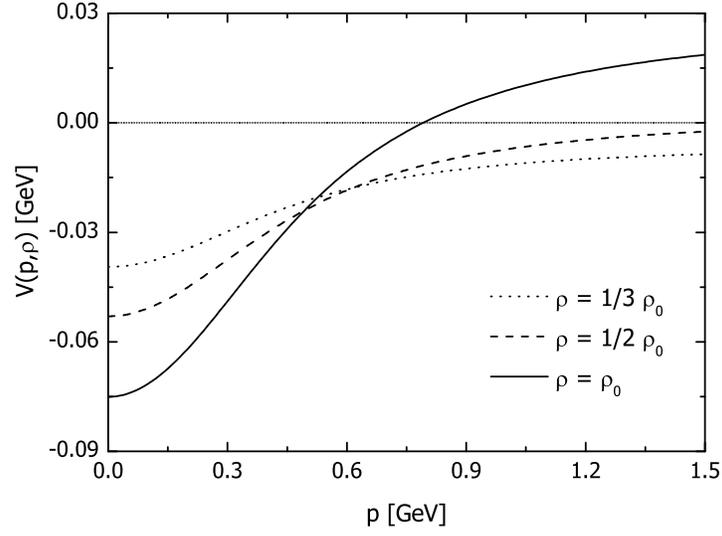}
\end{center}
\vspace{-0.5cm}
\caption{Momentum dependence of the potential $V$ with parameter set (M) given
in Tab. \protect\ref{tab:potpar} for different densities}. 
 \label{fig:potential}
\end{figure}

The single-particle energy of a nucleon in the local rest frame (LRF) of the
surrounding nuclear matter is given by
\begin{equation}
  \label{eq:potential1}
  \epsilon=\sqrt{(m_N-U_s)^2+p_{\textrm{LRF}}^2}+U_0\equiv
  \sqrt{m_N^2+p_{\textrm{LRF}}^2}+V
\end{equation}
with vector and scalar potentials $U_0$ and $U_s$.
For the nucleon, in the LRF the spatial components of the vector potential 
vanish in the mean-field approximation. In the non-relativistic limit, $V$ 
corresponds to the difference of scalar and vector potential.
For $V$ we use the density and momentum dependent non-relativistic mean-field
parametrization from Welke \emph{et al.} \cite{welke}
\begin{equation}
  \label{eq:potential2}
  V(\vec r,\vec p)=A{\rho\over\rho_0}+B\left({\rho\over\rho_0}\right)^{\tau}+
{8C\over\rho_0}\int{d^3p^\prime\over(2\pi)^3}{\Theta(p_F(\rho(r))-p^\prime)
\over 1+\left({\vec p-\vec p^\prime\over \Lambda}\right)^2}.
\end{equation}
In this equation $p_F=p_F(\rho)$ is the local Fermi momentum. We can work with
a choice 
of parameter sets $A,B,C,\tau,\Lambda$ determining the momentum dependence and 
stiffness of the nuclear matter equation of state (EOS) (cf. Table 
\ref{tab:potpar}).
Here, we use either a momentum independent hard EOS (H) or a momentum 
dependent medium EOS (M).
In Fig. \ref{fig:potential} we show the momentum dependence of the potential
$V$ for the parameter set (M) for different densities.
$V$ is transformed in the LRF into a scalar potential $S$ using
\begin{equation}
  \label{eq:potential3}
  \sqrt{(m_N+S)^2+p_{\textrm{LRF}}^2}=\sqrt{m_N^2+p_{\textrm{LRF}}^2}+V.
\end{equation}

We assume that the potential $V$ for the $S_{11}(1535)$ is also given by
Eq. (\ref{eq:potential2}), and the scalar resonance potential 
$S_R$ is obtained in the same way. The effective resonance mass is then 
given by
$$
  \mu_{\textrm{eff}}=\mu+S_R.
$$
Since no detailed calculations exist for the $S_{11}$, this is a reasonable
assumption.


\subsection{The elementary photon-nucleon reaction}

The main source of $\eta$ mesons is the elementary $\gamma N$ reaction.
It is known that in the threshold region the process $\gamma p\to \eta p$ 
is well described by the excitation of the $S_{11}$ resonance (see e.g.
\cite{penner}). 
The contribution coming from the two other resonances coupling to $N\eta$ 
are neglected, because the $\eta N$ branching ratio of the $S_{11}(1650)$ 
is very small and the $F_{17}(1990)$ is beyond the considered energy range.
Therefore, we use the following Breit-Wigner parametrization:
\begin{align}
  \label{eq:gamma_proton}
  \sigma_{\gamma p\to S_{11}\to X}&=\left({k_0\over k}\right)^2
   {s\Gamma_\gamma(\sqrt s)\Gamma_{S_{11}\to X}(\sqrt s)\over
   (s-M_{S_{11}}^2)^2+s\Gamma_{S_{11}\to X}^2(\sqrt s)} 
   {2m_N\over M_{S_{11}}\Gamma_0}\vert A_{1/2}^p\vert^2,\nonumber\\
  \sigma_{\gamma p\to S_{11}\to \eta p}&=\sigma_{\gamma p\to S_{11}}
     {\Gamma_{S_{11}\to \eta p}(\sqrt s)\over 
         \Gamma_{S_{11}\to X}(\sqrt s)}
\end{align}
with $\Gamma_\gamma=\Gamma_0\cdot k/k_0$ \cite{walker} and the
pole-mass decay width $\Gamma_0=0.151$ GeV.
The center of mass (cm) momentum $k$ of the $\gamma p$ pair
depends on the cm energy $\sqrt s$ (i.e. mass of the resonance), 
$k_0=k(M_{S_{11}})$ is the cm momentum taken at the pole mass of the $S_{11}$. 
The other (mass dependent) resonance widths $\Gamma_{S_{11}\to X}$ and
$\Gamma_{S_{11}\to\eta N}$ are parametrized as in
\cite{effe_dilep}. For the photocoupling 
helicity amplitude we use $A_{1/2}^p=0.109$ GeV$^{-1/2}$ \cite{krusche_eta}.
The cross section for the reaction $\gamma n\to S_{11}$ is obtained from 
(\ref{eq:gamma_proton}) by 
\begin{equation}
  \label{eq:gamma_neutron}
  \sigma_{\gamma n}={2\over 3}\sigma_{\gamma p},
\end{equation}
as suggested in \cite{krusche_deut}, corresponding to  
$A_{1/2}^n\sim 0.089$ GeV$^{-1/2}$. This value results from a resonance fit 
to eta photoproduction data on the deuteron and is connected to the value for
$A_{1/2}^p$ found in \cite{krusche_eta}. Therefore, we do not use the
(smaller) values found in other analyses (e.g. \cite{penner}).

\begin{figure}
\begin{center}
\includegraphics[width=11cm]{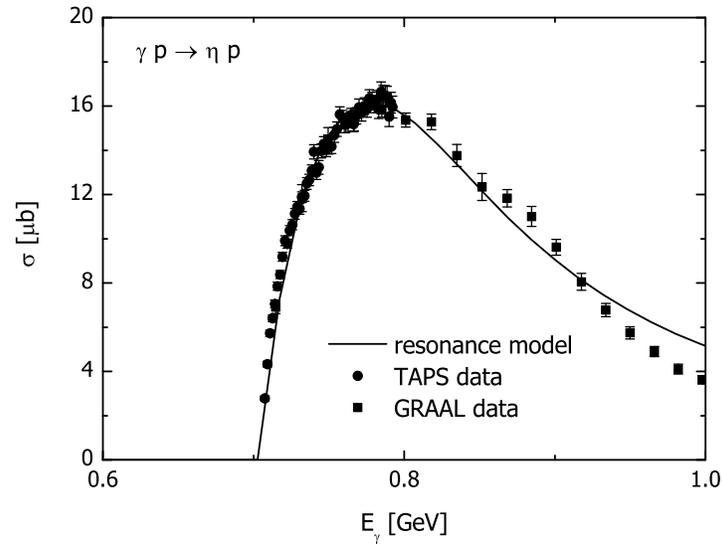}
\end{center}
\vspace{-0.5cm}
\caption{Parametrization of the elementary process $\gamma p\to\eta p$ 
according to Eq. (\protect\ref{eq:gamma_proton}). The data are taken
from \protect\cite{krusche_eta} and \protect\cite{graal_eta}.} 
 \label{fig:gamma_nucleon}
\end{figure}

In Fig. \ref{fig:gamma_nucleon} we show the cross section for the elementary
process $\gamma p\to\eta p$ in comparison with the data sets from 
\cite{krusche_eta,graal_eta}.
For energies up to $E_\gamma\sim 1$ GeV the agreement with the data is good.

Our model also contains other channels for the elementary $\gamma N$ 
interaction,
namely $\gamma N\to \pi N$, $\pi\pi N$, $P_{33}(1232)$, $D_{13}(1520)$,
$F_{15}(1680)$. These processes might also contribute to $\eta$ photoproduction
on nuclei via final state interactions (e.g. the reaction chain 
$\gamma N\to \pi N$, $\pi N\to S_{11}\to \eta N$). However, it was 
shown in \cite{effe_pion} that they are rather small.
The parametrization of these processes via resonance fits to experimental
pion photoproduction data is similar to Eq. (\ref{eq:gamma_proton}) and
described in Ref. \cite{effe_pion}.


\subsection{Final state interactions}

The $\eta$ final state interactions (FSI) are assumed to be mediated by
the re-excitation of resonances. In the energy range under consideration the 
$S_{11}(1535)$ clearly dominates. The cross sections for elastic and inelastic
$\eta N$ reactions therefore are similar to Eq. (\ref{eq:gamma_proton}) and 
strongly depend on the energy of the $\eta$:
\begin{equation}
  \label{eq:eta_fsi}
  \sigma_{\eta N\to S_{11}\to X}={4\pi\over p_{\eta N}^2}
 {s\Gamma_{S_{11}\to \eta N}(\sqrt s)\Gamma_{S_{11}\to X}(\sqrt s)
\over (s-M_{S_{11}}^2)^2+s\Gamma_{S_{11}\to X}^2(\sqrt s)}
\end{equation}
with the $\eta N$ cm momentum $p_{\eta N}$. $X$ contains the decay channels
$N\pi$, $N\eta$, $N\rho$, $N\sigma$ and $P_{11}(1440)\pi$. Among these,
$N\pi$ and $N\eta$ are the relevant channels (branching ratio $\sim 95\%$).
In the medium, the $S_{11}$ also undergoes FSI via the processes 
\begin{eqnarray}
  \label{eq:s11_fsi}
  N S_{11}  &\leftrightarrow& NN \nonumber\\
 N S_{11} &\to& N S_{11} \nonumber\\
N S_{11} &\leftrightarrow& N R,\ R\ne S_{11},
\end{eqnarray}
which give 
rise to a finite collision width. This will be discussed in 
Sec. \ref{sec:inmed_width}. Of course, the absorption of $S_{11}$ states in 
such reactions also contributes to the absorption of $\eta$ mesons.

Finally, the coupled-channel treatment of our BUU model also allows for 
contributions from side-feeding reactions such as $\pi N\to R\to \eta N$.
As already mentioned, the $\eta$ mesons managing to escape the nucleus
stem to a large extent from the elementary reaction chain 
$\gamma N \to S_{11} \to N\eta$.


\section{Calculation of the $\boldsymbol{S_{11}}$ in-medium width}
  \label{sec:inmed_width}

\begin{figure}
\begin{center}
\includegraphics[width=11cm]{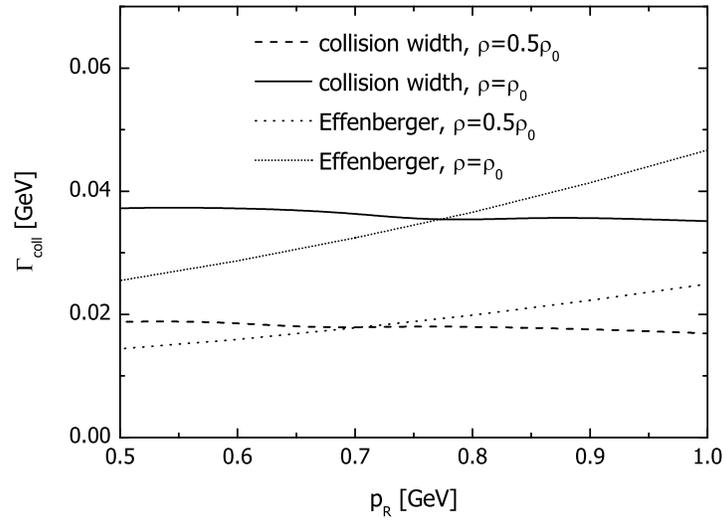}
\end{center}
\vspace{-0.5cm}
\caption{$S_{11}$ collision width at the pole mass as a function of the 
resonance momentum $p_R$ for different densities (dashed and solid curves).
Also shown are the results of Effenberger \emph{et al.} \protect\cite{effe_abs}
(dotted curves).}
 \label{fig:gcoll}
\end{figure}

In \cite{effe_abs} the collisional broadening of the resonances 
$P_{33}(1232)$, 
$D_{13}(1520)$, $S_{11}(1535)$ and $F_{15}(1680)$ was calculated from the 
collision rates resulting from processes like (\ref{eq:s11_fsi}):
\begin{align}
  \label{eq:collrate}
  \Sigma^>_{S_{11} N\to X}(E,p,\rho)={1\over 2E}&\intop_{}^{}
  {d^3 p_2\over (2\pi)^3}{d\mu_2\over 2E_2}{d^3 p_3\over (2\pi)^3}
  {d\mu_3\over 2E_3}{d^3 p_4\over (2\pi)^3}{d\mu_4\over 2E_4}
  (2\pi)^4\delta^4(p+p_2-p_3-p_4)\nonumber\\
  &\times \vert{\cal M}_{S_{11}N\to X}\vert^2\acal_2 f_2\acal_3 (1-f_3)\acal_4 (1-f_4)
\end{align}
with $X=NN$, $S_{11} N$ and $R N$. $f_2$ is the phase space density of the
incoming nucleons, $1-f_{3,4}$ are the Pauli blocking factors for the outgoing 
particles included in the final state $X$.
Note that such collision rates also appear in the collision integrals on the 
right-hand side of the BUU equations. Therefore, consistency between the collision 
widths and the processes explicitly included in the transport model is guaranteed.
The matrix elements appearing in the collision rates are identical to those 
contained
in the cross sections for the processes (\ref{eq:s11_fsi}).
Eq. (\ref{eq:collrate}) can be rearranged in the following way:
\begin{equation}
  \label{eq:ldt}
  \Sigma^>=\rho_N\langle v_{\textrm{rel}}^{\textrm{cm}}
    \sigma_{S_{11}N\to X}^{\textrm{cm}}(1-f_3)(1-f_4)\rangle_N,
\end{equation}
where the averaging is performed over the momentum distribution of the incoming
nucleon and $v_{\textrm{rel}}^{\textrm{cm}}$ is the relative velocity 
of the incoming $S_{11} N$ pair in the cm frame. We see that this expression is similar 
to what one obtains from the low density theorem
$\Gamma_{\textrm{coll}}=\rho\,v\,\sigma_{\textrm{tot}}$, leading to an increase of the
collision rates with momentum $p_R$, as seen in Fig. \ref{fig:gcoll}.
This way, for an on-shell resonance a broadening of about $35$ MeV at 
$\rho_0$ was found in \cite{effe_abs}.

\begin{figure}
\begin{center}
\includegraphics[width=14cm]{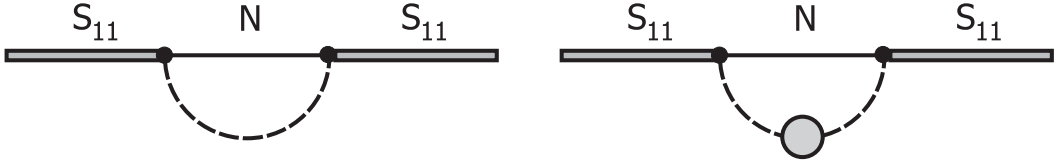}
\end{center}
\vspace{-0.5cm}
\caption{Feynman diagram for the self energy of the $S_{11}$ resonance in 
vacuum (left) and in nuclear matter (right). The dashed line represents the exchange of $\pi$, $\eta$ or $\rho$ mesons. By the blob in the meson line the full in-medium propagator for the mesons, which are dressed by the excitation of nucleon-hole and resonance-hole loops, is indicated.}
 \label{fig:selfen}
\end{figure}

It is well known from various calculations \cite{oset_delhole,post_rho1,lutz_kaon}, 
that the applicability of the low density theorem may be restricted to a regime of rather 
small densities. In order to overcome this problem we have developed a coupled channel 
analysis of the properties of $\pi$, $\rho$ and $\eta$ mesons as well as baryon 
resonances in 
nuclear matter \cite{post_rho1,post_rho2,post}. The resonance parameters are taken
 from the 
analysis of Manley \emph{et al.} \cite{manley}. The in-medium propagators of the 
mesons are 
determined from the excitation of nucleon-hole and resonance-hole loops. The resonance
self energy in 
nuclear matter $\Sigma_{\textrm{med}}$ is obtained by replacing the vacuum meson 
propagator with 
the in-medium one in the meson nucleon loops, see Fig. \ref{fig:selfen}. Of course, 
also the 
correction from Pauli blocking is included. The collisional broadening is given by:
\begin{equation}
 \label{eq:coll-width}
        \Gamma_{\textrm{coll}}(\sqrt{s},p) = 
        \frac{\textrm{Im}\,\Sigma_{\textrm{med}}(\sqrt{s},p)-
              \textrm{Im}\,\Sigma_{\textrm{Pauli}}(\sqrt s,p)}{\sqrt{s}},
\end{equation}
where $\textrm{Im}\,\Sigma_{\textrm{Pauli}}$ is the Pauli-blocked vacuum width.
By coupling the meson and resonance properties a self-consistency problem arises, 
which we 
solve iteratively. The physical interpretation is that higher iterations involve
 reactions on 
more than one nucleon. For example, going to the next order effectively takes into 
account 
3-body processes of the resonances. This is clearly beyond the low density theorem. 

Concerning the properties of the $S_{11}$ resonance, we find a net broadening of $35$ 
MeV at $\rho_0$ for an on-shell resonance. As can be seen in Fig. \ref{fig:gcoll}, 
in the momentum range of interest between 0.6 and 1 GeV this result is in absolute size
(relative to the total width at $\rho_0$) close to that of \cite{effe_abs}.
Effects from the dressing of $\pi$ and $\eta$ mesons are of the order of a few MeV, the 
only sizeable contribution to the collisional broadening coming from the $N\rho$ sector
(for details we refer the reader to \cite{post}). 
This is in qualitative agreement with the calculation of \cite{oset_eta}, where a 
somewhat larger value for the broadening was predicted.
The strong in-medium correction in the $N\rho$ channel is due to the coupling of the 
$D_{13}(1520)$ state to the $\rho$ meson, which moves spectral strength down to 
smaller invariant masses \cite{post_rho1,post_rho2} in the $\rho$ mass distribution. 
This way, the phase space available for the decay of a $S_{11}(1535)$ is enhanced.
The slightly different momentum behavior of the collision widths (\ref{eq:coll-width}) 
compared 
to the collision rates (\ref{eq:collrate}) is due to the resummation of particle-hole
loops in the meson propagators. This leads to a decrease of the $N\eta$ contribution and
an increase of the $N\rho$ contribution with increasing $p_R$, leaving the total width
$\Gamma_{\textrm{coll}}$ nearly constant. This behavior, however, cannot be extrapolated
to very small momenta $p_R<0.3$ GeV due to effects from Pauli blocking in the $N\eta$
channel.

We find that the main effect from the inclusion of higher order effects is a slight 
reduction 
of the broadening in the $N\rho$ channel. This is due to the strong broadening of the 
$D_{13}(1520)$ which leads to a moderate reduction of this state in the $\rho$ spectral 
function. Therefore, some $\rho$ spectral strength is moved up to larger invariant masses, 
and the decay of a $S_{11}$ to the $N\,\rho$ channel is somewhat suppressed.

We have also calculated the mass shift corresponding to the broadening of 
the $S_{11}$ by means of a dispersion analysis \cite{post}. As a result we do not 
observe any 
significant shift. This, of course, supports the hypothesis, that the mass shift as 
observed 
in the photo-nucleus data, see Fig. \ref{fig:gamma_nucleus}, is due to binding effects, 
see Section \ref{sec:results}. In the actual calculation, we therefore do not include
the dispersive mass shift.

The influence of the medium on the $S_{11}$ enters the calculations in two different
ways: The reactions displayed in (\ref{eq:s11_fsi}), which are directly 
connected with a collision rate via Eq. (\ref{eq:collrate}), are implemented explicitly 
in the model. Due to the 
fact that the results for the collisional width (Eq. (\ref{eq:coll-width})) are close to 
the collision rates, the consistency between the widths and these explicit processes is 
maintained. Furthermore, since the $S_{11}$ final states are not restricted
to the vacuum decay channels, the width $\Gamma_{S_{11}\to X}$ in Eqs.
(\ref{eq:gamma_proton}) and (\ref{eq:eta_fsi}) has to be substituted by the
full in-medium width.


\section{Results} \label{sec:results}

\begin{figure}
\begin{center}
\includegraphics[width=11cm]{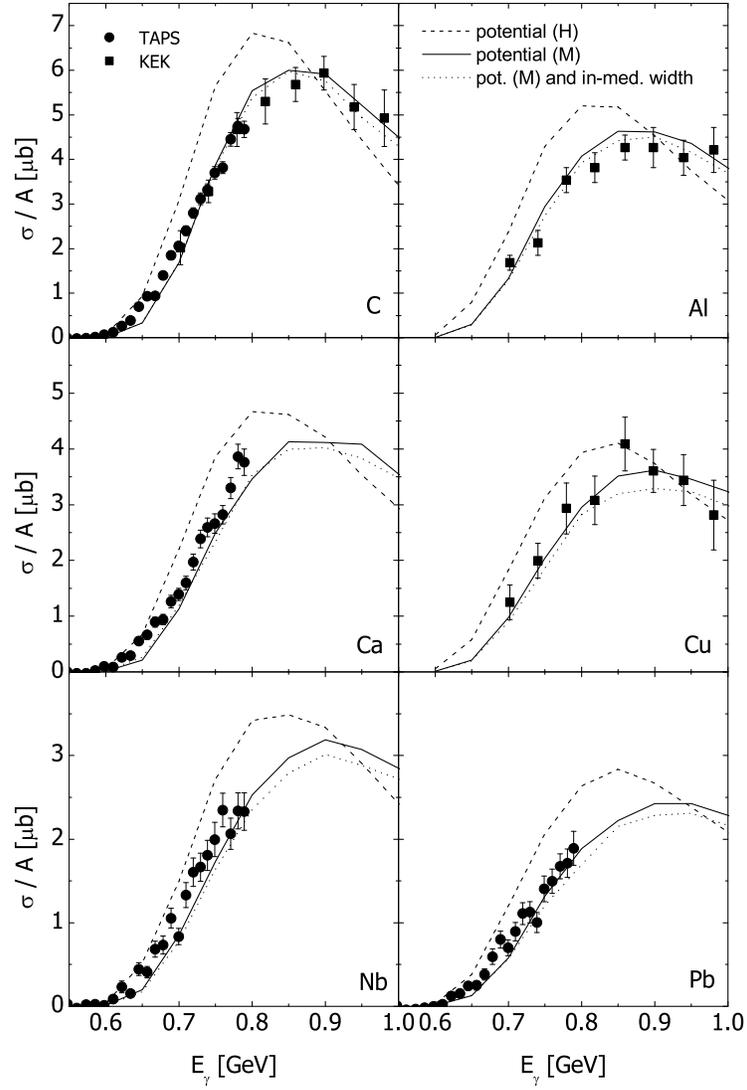}
\end{center}
\vspace{-0.5cm}
\caption{Results of the BUU model for the reaction $\gamma A\to\eta X$ on 
different nuclei. The dashed and solid lines correspond to the usage of the
potentials (H) and (M). The dotted lines include the medium-modified
width for the $S_{11}(1535)$ discussed in Sec. \ref{sec:inmed_width}.
The data are from \cite{roebig_eta} (circles) and 
\cite{yorita,yamazaki} (squares).}
 \label{fig:gamma_nucleus}
\end{figure}

In Fig. \ref{fig:gamma_nucleus} we show our results for the reaction $\gamma A\to \eta X$
for different nuclei in comparison with the TAPS and KEK data. The latter are obtained
by integration over $0^\circ\le \vartheta\le 90^\circ$, while the former data as well
as our calculations cover the full angular range. However, the main contribution stems from
angles smaller than $90^\circ$. Imposing this limitation on our calculations, deviations
smaller than 3\% are found.

The dashed and solid curves in each plot correspond to the two potentials (H) and (M) 
described in Sec. \ref{sec:buu_model}. The calculations include Fermi motion, Pauli blocking,
nuclear binding effects and FSI, but $S_{11}$ vacuum widths are used in the cross
sections (\ref{eq:gamma_proton}) and (\ref{eq:eta_fsi}). The dotted curves do
involve the full $S_{11}$ in-medium width in the cross sections and therefore show 
the total influence of collisional broadening.

The curves (M) are shifted by about 50-70 MeV compared to the curves (H).
This can be understood by taking a look at the elementary reaction
$\gamma N\to S_{11}(1535)$. The momentum independent potential (H) is 
identical both for incoming nucleon and outgoing resonance. 
The resonance mass is obtained from the cm energy $\sqrt s$ (i.e. effective resonance 
mass) by subtracting the resonance potential:
$$
  \mu_{\textrm{H}}=\sqrt s - S_{\textrm{H}}.
$$
In the case of the momentum dependent potential (M) we obtain a different result. For 
photon energies of about 800 MeV the outgoing resonance has also a momentum of
about 800 MeV. For such momentum values, the potential (M) is almost zero, see Fig. 
\ref{fig:potential}.
Therefore we get for the resonance mass $\mu_{\textrm{M}}=\sqrt s$, which is 
is smaller compared to $\mu_{\textrm{H}}$:
$$
  \mu_{\textrm{M}}=\sqrt s=\mu_{\textrm{H}}-\vert S_{\textrm{H}}\vert.
$$
Hence the peak maxima in both scenarios are shifted relatively to each other
by roughly $\vert S_{\textrm{H}}\vert$, which is approximately 50-70 MeV.

As one can see, the curves (H) overestimate the TAPS data in the threshold
region and exhibit a maximum shifted towards lower energies 
with respect to the peak maximum suggested by the KEK data. In contrast, the curves (M) only
slightly underestimate the TAPS data, but are in very good
agreement with the KEK data, especially the location of the peak maximum is well
reproduced. 

\begin{figure}
\begin{center}
\includegraphics[width=11cm]{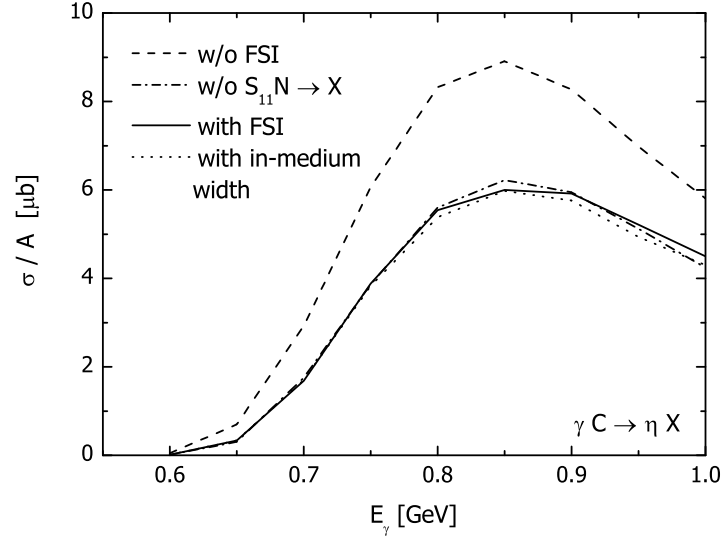}
\end{center}
\vspace{-0.5cm}
\caption{Influence of medium modifications on $\eta$ photoproduction
on Carbon with potential (M). The dashed curve shows the result without
FSI, the dashed-dotted includes FSI except for the $S_{11}$ collision
reactions. The solid curve includes all FSI and vacuum widths in the 
cross section for the $S_{11}$ excitation
and corresponds to the
solid line in Fig. \protect\ref{fig:gamma_nucleus}. The dotted curve
includes full in-medium widths in these cross sections and 
coincides with the dotted curve in Fig. \protect\ref{fig:gamma_nucleus}.}
 \label{fig:carbon_coll}
\end{figure}

The influence of the collisional broadening of the $S_{11}$ has only little influence. As can be seen from
the dotted curves, which visualize the effect of the medium modification via
the cross sections (\ref{eq:gamma_proton}) and (\ref{eq:eta_fsi}). 
In order to completely analyze the influence of collisional broadening, we show in
Fig. \ref{fig:carbon_coll} four different szenarios:
The dashed curve shows the calculation without any FSI. The dashed-dotted curve includes
FSI except for the $S_{11}$ FSI in (\ref{eq:s11_fsi}) and therefore
shows the influence of the direct $\eta$ absorption via resonance re-excitation.
Both calculations use vacuum widths in the cross sections (\ref{eq:gamma_proton}) and
(\ref{eq:eta_fsi}). The solid curves include all FSI and corresponds to the  solid
curves in Fig. \ref{fig:gamma_nucleus}. The dotted curve also includes the 
full in-medium widths in the cross sections and corresponds to the dotted curves in
Fig. \ref{fig:gamma_nucleus}. It is seen that the main effect comes from the direct $\eta$
absorption via $\eta N\to R$, whereas the influence of in-medium effects concerning the 
$S_{11}$ is rather small. This is not surprising, because the mean free path $\lambda$ of 
a $S_{11}$ at $\rho_0$ is about 3 fm -- compared to $\lambda\lesssim 1$ fm for the $\eta$
-- whereas the RMS radius of C is about 2.5 fm. 
This is in contrast to the results of Yorita \emph{et al.} \cite{yorita}, who find
a strong effect from the $S_{11}$ FSI.

\begin{figure}
\begin{center}
\includegraphics[width=11cm]{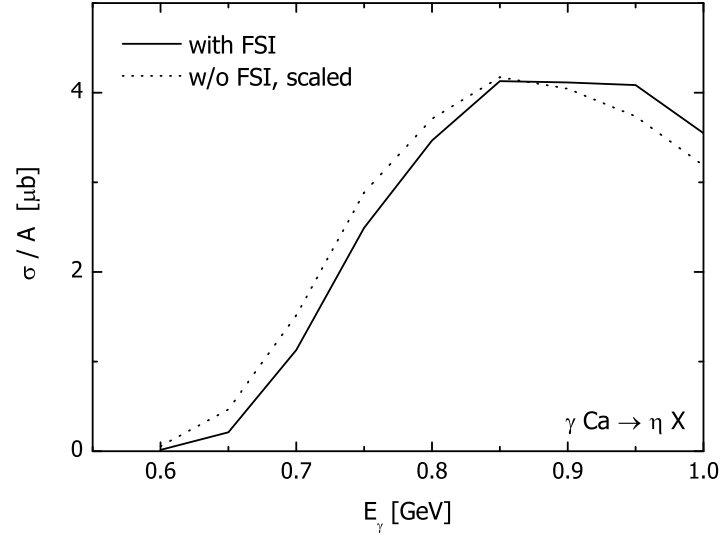}
\end{center}
\vspace{-0.5cm}
\caption{Influence of the FSI on the cross section $\gamma \textrm{Ca}\to\eta X$.
The solid curve shows the result with FSI, while the 
dotted line is the result without FSI divided by a
factor of 1.9. All curves include the potential (M). In-medium 
modifications of the $S_{11}$ width are neglected.}
 \label{fig:ca_fsi}
\end{figure}

In Fig. \ref{fig:ca_fsi} we show the effect of the FSI in the case 
of Calcium. The solid line includes FSI. The dotted curve is the result 
without FSI divided my a constant
factor of 1.9. In both calculations we used the potential (M) and neglected the medium 
modification of the $S_{11}$ widths. It is clearly visible that the absorption depends on 
energy and therefore the assumption of a constant absorption factor made in 
\cite{maruyama} is unrealistic. This result also holds for other nuclei.
On the other hand, the issues raised in \cite{maruyama} concerning the
peak position can be verified. The claim there was that the peak position can
only be described by assuming vanishing scalar and vector potentials for the 
$S_{11}(1535)$. This scenario is close to what we get with the momentum
dependent potential (M) for the $S_{11}$, which -- as mentioned above -- 
nearly vanishes in the kinematical regime under investigation.


\section{Summary}

We have calculated $\eta$ photoproduction on several nuclei in the energy
region dominated by the $S_{11}(1535)$ resonance. Starting from a parametrization
of the data for the elementary $\gamma N\to\eta N$ reaction, we applied a BUU
transport model to account for final state interactions.
The calculations including a momentum dependent potential for the nucleons and the 
$S_{11}$ reproduce the available data well. In the threshold region a slight 
underestimation of the data is observed. The calculations involving a momentum 
independent potential cleary overestimate the data.

The collisional width of the $S_{11}$, calculated within a realistic self-consistent 
resonance-hole model 
\cite{post}, was found to be close to the collision rates from \cite{effe_abs}
and was included in the transport calculations.
The influence of collisional reactions of the $S_{11}$ on the cross section were found 
to be small, in contrast to the QMD calculations of Yorita \emph{et al.} \cite{yorita}.
The main difference between the result without FSI and the full calculation can be
attributed to the absorption of the $\eta$ in the FSI e.g. via 
$\eta N\to S_{11}\to \pi N$.
Finally, we have shown that the absorption reactions cannot be mimicked by applying
a constant factor to the results without FSI.

\section*{ACKNOWLEDGMENTS}

This work was supported by DFG.


\newpage

\begin{table}
\caption{\label{tab:potpar} Parameters for the mean-field potential $V$.}
\begin{ruledtabular}
\begin{tabular}{lcccccc}
   & incompressibility [MeV] & $A$ [MeV] & $B$ [MeV] & $C$ [MeV] & $\tau$ & 
$\Lambda$ [fm$^{-1}$] \\
 \hline
 H & 380 &-124.3 & 71.0 & 0. & 2.0 & - \\
 M & 290 &-29.3 & 57.2 & -63.5 & 1.76 & 2.13 \\
\end{tabular}
\end{ruledtabular}
\end{table}

\end{document}